\documentclass[12pt]{article}
\usepackage{epsfig}
\usepackage{rotating}
\usepackage{url}
\newcommand{\be}{\begin{equation}}
\newcommand{\ee}{\end{equation}}
\newcommand{\ba}{\begin{eqnarray}}
\newcommand{\ea}{\end{eqnarray}}

\begin{document}

\begin{center}
\Large{\textbf{Nurturing Breakthroughs: \\Lessons from Complexity Theory}}\\~\\
\normalsize{D. Sornette} 
\end{center}
\begin{center}
\normalsize{
Department of Management, Technology and Economics,
ETH Zurich\\ 
Kreuzplatz 5, CH-8032 Zurich, Switzerland
}\\
\normalsize{\today}\\~\\
\end{center}

\begin{abstract}
A general theory of innovation and progress in
human society is outlined, based on the combat between two opposite forces (conservatism/inertia
and speculative herding ``bubble" behavior).  We contend that
human affairs are characterized by ubiquitous ``bubbles'',
which involve huge risks which would not otherwise be taken using standard cost/benefit analysis. Bubbles result from
self-reinforcing positive feedbacks. This leads to explore uncharted
territories and niches whose rare successes lead to
extraordinary discoveries and provide the base for
the observed accelerating development of technology and of the economy.
But the returns are very heterogeneous, very risky
and may not occur. In other words, bubbles, which 
are characteristic definitions of human activity,  allow huge risks to get huge returns over
large scales. 
We outline some underlying mathematical structure and a few results involving
positive feedbacks, emergence, heavy-tailed power laws,
outliers/kings/black swans, the problem of predictability and
the illusion of control, as well as some policy implications.
 
\end{abstract}

\vskip 1cm

Invited talk at the workshop Trans-disciplinary Research Agenda for Societal Dynamics (\url{http://www.uni-lj.si/trasd in Ljubljana}), organized by Professor J. Rogers Hollingsworth, Dr. Karl H. Mueller, Professor Ivan Svetlik (vice rector of the University of Ljublana), 24 - 25 May 2007, Ljubljana, Slovenia
 \vskip 1cm

In order to address the question, ``how can complexity science support policy
decision making?'', it seems to me that one needs first to embrace the larger
question of what is driving a human being, a human groups or a human society.
Only then can policy decision making identify and aim correctly its targets.
My perception is that these questions can now be addressed with a fresh perspective,
informed and inspired by science. Indeed, the
scientific enterprise is now offering insights on many of the questions in social
sciences that were before only the realm of normative approaches or philosophical discourses.
Via novel experimental procedures,
new technology and new conceptual insights, science can now help address
from an evidence-based approach such questions as:
what is happiness? What is driving human beings? Why do we cooperate?

My discussion is organized in four short parts. 
\begin{enumerate}
\item The first one recognizes that most systems
are punctuated by rare but large events which often dominate their organization.
The progress of science and technology is no exception, as 
innovations, discoveries, blockbusters are exceptional events in their impact.
This statement can be quantified by heavy-tailed distributions.
\item Complementing the first part which emphasizes the statistics of the impacts
of innovations, the second part focuses on a classification of their time dynamics,
by presenting a simple classification of the dynamics of complex
systems in terms of the interplay between endogenous and exogenous shocks. 
This allows us to separate the impact of external influences from the role
played by the constitution of the inner structure in understanding and predicting
blockbusters. This is applied to different critical events in economics and the social
sciences taken as proxies of innovation dynamics: social unrest shocks, internal downloads, dialog in email traffic, the dynamics
of commercial sale in response to advertisement or to word-of-mouth, volatility clustering
and shocks in financial markets, dynamics of exploits 
and patches following disclosures of software vulnerability, movie blockbusters, the
dynamics of YouTube video sharing, and so on.
\item The third part documents the phenomenon that I coin ``breakdown of the human 
Galilean invariance principle'', namely that humans  being bored by steady state
tend to act to develop intermittent accelerating outcomes.
As a consequence of their individual actions aggregated at the collective level, 
one can observe super-exponential acceleration of their processes punctuated
by corrections and crashes. The mathematical description of these processes
emphasizes the importance of positive feedbacks. Based on this idea,
I propose that ``bubbles'' are generic
results of collective human activities and that they seem to be not only inherently
associated with human societies but are also a vehicle of giant leaps in progress.
\item Lastly, the question of control and management of complex systems
is alluded to by stressing the phenomenon of  ``illusion of control'' and its
consequence for practical policies.
\end{enumerate}
The last section concludes.

\section{Heavy-tail distribution of breakthroughs and blockbusters}

Many studies report evidences of positive economic benefits derived from basic research
[Martin et al., 1996; NAS, 1997]. In certain areas such as biotechnology, semi-conductor physics, 
optical communications [Ehrenreich, 1995], the impact of basic 
research is direct while, in other disciplines, the path from discovery to applications
is full of surprises. As a consequence, there are persistent uncertainties in the quantification of 
the exact economic returns of public expenditure on basic research. This gives little help
to policy makers trying to determine what should be the level of funding.

Some industries, such as the pharmaceutical and movie industries, are characterized by the occurrence of ``block-busters,'' i.e. remarkably successful products with exceptional sales much larger than the typical product. Determining how exceptional are these block-busters is an important question for firm strategy and economies of scale. Taking the movie industry as a proxy because the data is unambiguous, plentiful and of good quality, Sornette and Zajdenweber [1999] have shown that the distribution of gross revenues of Hollywood movies from theatres from top box office 100 is stable over twenty years and is well-described by a power law distribution with exponent $\alpha$ approximately equal to $1.5$, such that the mean exists but not the variance. Grabowski and Vernon [1990, 1994] constructed a discounted present value per new chemical entity and divided the drugs in decile in descending order, leading to a value distribution compatible with a power law distribution of the tail with exponent approximately equal to $2/3$ [Sornette, 2002a; and unpublished], for which neither the mean not the variance exist. Scherer [1998] has studied the distribution of royalties from U.S. University patent portfolios, of the quasi-rents from marketed pharmaceutical entities and the stock market returns from three large samples of high-technology venture start-ups. The tails of the distributions are again compatible with a power law distribution with exponent less than $1$ but there is a noticeable curvature when going to small returns. D. Harhoff and F.M. Scherer  (private communication) have studied a sample of approximately 800 high-value patents and find that the distribution most closely approximates a log normal while the power law hypothesis is strongly rejected when using the whole sample. The issue whether the extreme tail of the distribution of returns from innovation is asymptotically a power law with a small exponent is thus a delicate statistical problem. This is not specific to this domain of  application, see for instance Sornette [2004].

Sornette and Zajdenweber [1999] have suggested that these uncertainties 
on what should be the relevant policy on research support have a 
fundamental origin to be found in the interplay between
the intrinsic ``fat tail'' power law nature of the distribution of economic returns of innovations, characterized by
a mathematically diverging variance, and the stochastic character of 
discovery rates. In the regime where the cumulative
economic wealth derived from research is expected to exhibit a long-term positive trend, they show 
that strong fluctuations blur out significantly the short-time scales\,:
a few major unpredictable innovations may provide a finite 
fraction of the total creation of wealth. In such a scenario, 
any attempt to assess the economic impact of research over a finite time horizon
encompassing only a small number of major discoveries is bound to be at best
unreliable and at worst misleading.

In the Kuhnian view of how science works [Kuhn, 1970], periods of ``normal science'' 
are interrupted by revolutions. If discovery ``sizes'' are indeed distributed according
to a power law distribution, it is natural to wonder
if Kuhn was only half-correct\,: it seems possible that there is no such thing 
as ``normal science', and that science instead evolves through a 
succession of ``revolutions'' of all sizes. This idea has been put forward in
another context by M. Buchanan [1996]. Accordingly, history only takes 
notice of the really huge ``revolutions'' --quantum theory and relativity, for 
example-- even though there are  less significant others going on 
all the time. As a signature of this idea, 
there should be some kind of Gutenberg-Richter law for ideas --a power 
law distribution of their impact, as found by Redner [1998] and
Dieks and Chang [1976] for the impacts of scientific 
publications.

This suggests to bring the problem of research economic benefits into 
the growing basket of natural and societal processes characterized by extreme behavior.
They range from large natural catastrophes such as volcanic
eruptions, hurricanes and tornadoes, landslides, avalanches, lightning strikes,
catastrophic events of environmental degradation,
to the failure of engineering structures, social
unrest leading to large-scale strikes and upheaval, economic drawdowns on national
and global scales, regional power blackouts, traffic gridlock, diseases and
epidemics, etc. These phenomena are extreme events that occur rarely,
albeit with extraordinary impact, and are thus completely under-sampled and thus 
poorly constrained. They seem to result from self-organising systems 
which develop similar patterns
over many scales, from the very small to the very large. There is an urgency
to assimilate in our culture and policy that we are embedded in extreme phenomena.
Our overall sense of continuity, safety and confort may just be an illusion stemming from
our myopic view. Let us unleash the battle of giants between extraordinary discoveries and
extreme catastrophes.

\section{Interplay between Endogenous and Exogenous shocks (endo-exo): from commercial sales to happiness}

\subsection{Motivations}

Self-organized criticality, and more generally, complex system theory
contend that out-of-equilibrium slowly driven systems with threshold
dynamics relax through a hierarchy of avalanches of all sizes.
Accordingly, extreme events are seen to be endogenous [Bak and Paczuski, 1995; Bak, 1996] in contrast with
previous prevailing views. In addition, the preparation processes
before large avalanches are almost
undistinguishable from those before small avalanches, making the prediction
of the former basically impossible (see [Sornette, 2002b] for a discussion).
But, how can one assert with 100\% confidence
that a given extreme event is really due to an endogenous
self-organization of the system, rather than to the response to an
external shock? Most natural and social systems are indeed continuously
subjected to external stimulations, noises, shocks, sollications,
forcing, which can widely vary in amplitude. It is thus not clear a
priori if a given large event is due to a strong exogenous shock, to the
internal dynamics of the system organizing in response
to the continuous flow of small sollicitations, or maybe to a combination of both.
Adressing this question is fundamental for understanding the relative importance of
self-organization versus external forcing in complex systems and for the 
understanding and prediction of crises.

This leads to two questions:
\begin{enumerate}
\item Are there distinguishing properties that
characterize endogenous versus exogenous shocks?

\item  What are the relationships between endogenous and exogenous shocks?
\end{enumerate}
Actually, the second question has a long tradition in physics. It is 
at the basis of the interrogations that scientists perform on the enormously
varied systems they study. The idea is simple: subject the system to 
a perturbation, a ``kick'' of some sort, and measure its response as a function
of time, of the nature of the sollicitations and of the various environmental
factors that can be controlled. In physical systems at the thermodynamic
equilibrium, the answer is known under the name of the theorem of fluctuation-dissipation, 
sometimes also refered to as the theorem of fluctuation-susceptibility [Stratonovich, 1992]
In a nutshell, this theorem relates quantitatively in a very precise way
the response of the system to an instantaneous kick (exogeneous) to the correlation function
of its spontaneous fluctuations (endogenous). An early example of this relationship is
found in Einstein's relation between the diffusion coefficient $D$ of 
a particle in a fluid subjected to the chaotic collisions of the fluid molecules
and the coefficient $\eta$ of viscosity of the fluid [Einstein, 1905; 1956]. 
The coefficient $\eta$ controls the drag,
i.e., response of the particle velocity when subjected to an exogenous force impulse.
The coefficient $D$ can be shown to be a direct measure of the (integral of the) 
correlation function of the spontaneous (endogenous) fluctuations of the particle velocity.

In out-of-equilibrium systems, the existence of a relationship between 
the response function to external kicks and
spontaneous internal fluctuations is not settled [Ruelle, 2004].
In many complex systems, this question amounts to
distinguishing between endogeneity and exogeneity and is important
for understanding the relative effects of self-organization versus
external impacts. This is difficult in most physical systems because
externally imposed perturbations may lie outside the complex attractor
which itself may exhibit bifurcations. Therefore, observable perturbations
are often misclassified. 

It is thus interesting to study other systems, in which the dividing line
between endogenous and exogenous shocks may be clearer in the hope that it
would lead to insight about complex physical systems. The investigations
of the two questions above may also bring new understanding of these systems.
The systems to which the endogenous-exogenous question (which we will refer to as ``endo-exo''
for short) is relevant include the following:
\begin{itemize}
\item Biological extinctions such as the
Cretaceous/Tertiary KT boundary (meteorite versus extreme volcanic
activity (Deccan traps) versus self-organized critical extinction cascades), 

\item immune system deficiencies (external viral/bacterial infections versus
internal cascades of regulatory breakdowns), 

\item cognition and brain learning processes (role of
external inputs versus internal self-organization and
reinforcements),

\item discoveries (serendipity versus the outcome of slow endogenous maturation
processes), 

\item commercial successes (progressive reputation cascade versus the result of a well
orchestrated advertisement),

\item financial crashes (external shocks versus self-organized instability),

\item intermittent bursts of financial volatility (external shocks versus
cumulative effects of news in a long-memory system),

\item the aviation industry
recession (9/11/2001 terrorist attack versus structural endogenous problems),

\item social unrests (triggering factor or rotting of social tissue),

\item recovery after wars (internally generated
(civil wars) versus imported from the outside) and so on.
\end{itemize}

It is interesting to mention that the question of exogenous versus endogenous forcing has been
hotly debated in economics for decades. A prominent example is the
theory of Schumpeter on the importance of technological discontinuities in economic history.
Schumpeter argued that ``evolution is
lopsided, discontinuous, disharmonious by nature... studded with violent outbursts and
catastrophes... more like a series of explosions than a gentle, though incessant,
transformation'' [Schumpeter, 1939]. Endogeneity versus exogeneity is also
paramount in economic growth theory [Romer, 1996].

\subsection{Epidemic model of social interactions by word-of-mouth}

Our approach to the endo-exo question is based on epidemic cascade models
of social interactions (see Sornette, 2005 and http://www.er.ethz.ch/essays/origins for reviews).
Let us consider an observable characterizing the activity of humans
within a given social network of interactions. This activity can 
be the rate of visits or downloads on an internet website, the sales of a book
or the number of newpaper articles on a given subject. 
We envision that the instantaneous activity
results from a combination of external forces such as
news and advertisement, and of social influences in
which each past active individual may impregnate other individuals in
her network of acquaintances with the desire to act. This
impact of an active individual onto other humans is not instantaneous as
people react at a variety of time scales. The time delays capture
the time interval between social encounters, the maturation of 
the decision process which can be influenced by mood, sentiments, and many
other factors and the availabilty and capacity to implement the decision.
The contacts and exchanges between humans lead to information cascades.

This leads to identify a critical parameter, the branching ratio $n$, which controls
the propensity for news, shocks, changes to propagate and percolate within
the social network. When $n<1$, the social network is sub-critical and 
local changes remain localized. When $n$ reaches $1$, the system is critical,
characterized by power laws in the distribution of group sizes impregnated
by a given change. For $n>1$, the network is super-critical as local changes
have a finite probability to propagate and invade the entire network. The
regime $n \geq 1$ is the one of interest for marketing campaigns, for policy 
targets, more generally, for any action that aims at a maximum impact with a
minimal cost. Typically, an external input of amplitude $S$ is amplified by the
network effect of word-of-mouth epidemics by the factor $1/(1-n)$ (for $n<1$), 
suggesting that policy making and marketing campaigns should not only 
optimize their action $S$ but also target mature and receptive networks characterized
by a branching ratio close to, equal to or even larger than $1$.

These ideas and quantitative results are relevant to social unrest shocks, internal downloads, dialog in email traffic, the dynamics of commercial sale in response to advertisement or to word-of-mouth, volatility clustering and shocks in financial markets, dynamics of exploits 
and patches following disclosures of software vulnerability, movie blockbusters, the
dynamics of YouTube video sharing, and so on. 

\subsection{A conjecture on Happiness}

Recent works by Kahneman and others show that humans have a kind
of reference point for happiness [Reichhardt, 2006]. This reference point may be
different from one human being to the next. From two kinds of
interviews, researchers have documented that humans are subjected
to burst of happiness or despair which then relax after some time
to their previous happiness level, similar to the relaxation of
the response function due to an exogenous shock. Someone who gains
a big lottery or someone who suffers a major accident leading to
paralysis for instance will have a large instantaneous perturbation in their
level of happiness, but remarkably both will return after some time
and some adjustment to basically their previous level of 
happiness. 

This suggests that happiness and well-being in humans can 
be approached by the generic endo-exo approach outlined in the previous
section. The guideline
offered by this insight is that we need to understand what are
the individual, cultural and societal variables that may control 
the  ``criticality'' parameter $n$ of our response to the incessant flux
bathing our life, so that, under a constant background of  
``happiness sources'' $S$, the overall happiness level is amplified to 
$S/(1-n)$ by internal and social cascades.  
This suggests to emphasize the policy impact on $n$ (our internal state) more
than on $S$, the external sources of improvements.

\section{Bubbles everywhere}

\subsection{Breakdown of ``Galilean invariance principle'' in human psychology}

I propose the idea that the collective dynamics of human affairs exhibit
a breakdown of ``Galilean invariance principle''. In physics, 
Galilean Invariance is a principle of relativity which states that the fundamental laws of physics are the same in all inertial frames. Its direct consequence is that deviations of a body from constant 
velocity can only occur upon the application of a force. The analogy is that 
constant velocity or rate of change is boring to human beings, 
who tend to act to develop intermittent accelerating outcomes.
As a consequence of their individual actions aggregated at the collective level, 
one can observe super-exponential acceleration of their processes punctuated
by corrections and crashes. I take as evidence of this the example of 
stock market bubbles and crashes. The perhaps most striking illustration
is the market of Hong Kong, arguably until 1997 the freer market in the world
with complete flexibility in the mobility of capital and its investment use. 
While the Hang Seng Hong Kong index from 1979 to 1997 is characterized
by an approximately constant annual return of $14\%$, this average return
is a very poor description of what was actually occurring in the market: either it
was super-exponential accelerating, or it was crashing. This observation has been
described and generalized to many bubbles and crashes in different parts of 
the world and at many epochs [see Sornette, 2003a,b and references therein]. 

The underlying mechanism
involves positive feedbacks on prices, i.e., to the fact that, conditioned
on the observation that the market has recently moved 
up (respectively down), this makes it more probable to keep it moving up 
(respectively down) in an amplified move, so that a large cumulative move ensues. 
The concept of ``positive feedbacks'' has a long
history in economics and is related to the idea of 
``increasing returns''-- which says that goods become cheaper the more of them
 are produced (and the closely related idea that some products, like fax
 machines, become more useful the more people use them). 
 ``Positive feedback'' is the opposite of ``negative feedback'', 
 a concept well-known for instance in population dynamics: the larger the population
 of rabbits in a valley, the less they have grass per rabbit. If the population
 grows too much, they will eventually starve, slowing down their reproduction rate
 which thus reduces their population at a later time.
 Thus negative feedback means that the higher the population, the 
 slower the growth rate, leading to a spontaneous regulation of 
 the population size; negative feedbacks thus tend to regulate growth
 towards an equilibrium.

 In contrast, positive feedback asserts that the 
 higher the price or the price return in the recent past, the higher will be
 the price growth in the future. Positive feedbacks, when unchecked, can produce
 runaways until the deviation from equilibrium is so large that other 
 effects can be abruptly triggered and lead to rupture or crashes.
 The positive feedback
leads to speculative trends which may
dominate over fundamental beliefs and which make the system
increasingly susceptible to any exogenous shock, thus eventually precipitating a
crash. There are many mechanisms in the stock
market and in the behavior of investors which may lead to positive feedbacks.
They can be roughly divided into two classes.
\begin{itemize}
\item Technical and rational mechanisms for positive feedbacks:
	\begin{enumerate}
	\item option hedging,
	\item insurance portfolio strategies,
	\item trend following investment strategies,
	\item asymmetric information on hedging strategies.
	\end{enumerate}
\item Behavioral mechanism for positive feedbacks based on imitation:
	\begin{enumerate}
	\item it is rational to imitate,
	\item imitation is the highest cognitive task,
	\item we mostly learn by imitation,
	\item cultural imitation and ``conventions.''
	\end{enumerate}
\end{itemize}
Sornette [2003a,b] present detailed documentation and arguments 
emphasizing the positive feedbacks by imitation, which
is also known as the  ``herd'' or ``crowd'' effect.
 
\subsection{Positive effect of ubiquitous ``bubbles'' in human affairs}
 
I propose to generalize the observation that financial markets exhibit alternating regimes 
of over-enthusiasm leading to bubbles followed by phases of
consolidation, bearish trends or even crashes. 
In finance and economics, the term ``bubble'' refers to a situation in which excessive public expectations of future price increases cause prices to be temporarily elevated without justification from fundamental valuation. I extend this definition to human affairs as follows.

{\it Definition of a ``bubble'' in human affairs}: a ``bubble'' occurs when 
excessive public/political expectations of positive outcomes cause over-enthusiasm and unreasonable investment and efforts.

During bubbles, people take inordinate risks that would not otherwise be justified by standard cost-benefit and portfolio analysis. Instead, people rationalize their risk-taking behavior by new models of net-present-value, such as witnessed during the new economy bubbles of IT and Internet companies that culminated in 2000: the new accounting method over-emphasized the ``real option'' value of companies associated with the new niches that they were opening. I want to emphasize the role of bubbles in human affairs because they seem inherently associated with the innovation process and the creation of new technology. Bubbles lead to a lot of destruction of value but also to the exploration and discovery of exceptional niches. Only during these times do people dare explore new opportunities, many of them unreasonable and hopeless, with rare emergences of great lucky outcomes. This is the wild risk regime of extremely heavy tails in the distribution of economic returns on investments. I envision this mechanism as the leading one controlling the appearance of disruptive innovations and major advances. In a word, bubbles (collective over-enthusiasm) seem a necessary evil to foster our collective attitude towards risk taking and break the stalemate of society resulting from its tendency towards risk avoidance. Bubbles are times of self-organized self-excited auto-catalytic amplification of 
risks that allow the exploration of new niches. I contend that society needs 
bubbles because the bubbles lead to a very risky behavior that lead to great
potential returns. In absence of bubble psychology, no large risks are taken and no large
return can derive leading to stagnation. 

To illustrate this hypothesis, many examples can be put forward. Here, I discuss briefly
only a few salient cases.
\begin{itemize}
\item {\bf Great boom of railway shares in Britain in the 1840s.} Consider the great boom in railway shares in Britain in the 1840s. The boom and collapse not only depleted the wealth of many
individuals, but cut briefly into the capital available for normal trade and
finance. The overbuilding of
railroads meant that few could earn back a return commensurate with the capital
put into them.

Yet,  Britain ended up with an extensive railway system ahead of other
industrializing nations. Even if the building was done inefficiently, the gains
to the economy were rather large.

\item {\bf Appolo program.} On the fifth flight, Appolo 11 landed on the moon, July 20, 1969; Armstrong and Aldrin became the first humans to land and walk, at an estimated cost of 20 to 25 billion (of 1969 US dollars). At this time, what were the anticipations for the following 30 years, at the horizon of the new millenia? Great expectations included to put the Moon and Mars in ``mankind's sphere of economic influence'' (to use a phrase later chiseled by  Presidential Science Advisor Jack Marburger). In the late 1960s, it was thought that permanent bases on the Moon would be routinely operated and that mankind would have already landed on Mars and beyond.

What is the state of space exploration and conquest in 2007? Would mankind be able to land again on the Moon and at what cost?  Personally, as a teenager at the time of the historical landing who has become a grown-up over these more than 30 years, and as a scientist having contributed to 
space science [Anifrani et al., 1995; 1999; Maveyraud et al., 2001; Le Floc'h and Sornette, 2003], 
I find the present situation on space exploration
quite disappointing and depressing. It is clear that the expectations have been unfulfilled and that there are still major obstacles to overcome: protection of humans from cosmic rays, medical problems
appearing in absence of gravity, reliability of spacecrafts,  propulsion efficiency to cite a few.

It is clear that, as vividly expressed by NASA administrator Michael Griffin [2007], space exploration is perhaps the best incarnation of the ``real reasons'' for taking risks in unreasonable (to standard cost/benefit analysis) endeavors. The ``real reasons'' that go beyond reason perhaps help define human beings. They include the enthusiasm for new things, the wonder and awe of discoveries, the challenges of competition, the taking of hard challenges... for the sake of the challenge.

\item {\bf Human Genome project.} ``The ultimate goal of this initiative is to understand the human genome'' (US Department of Energy, 1984, 1986). A 'rough draft' of the genome was finished in 2000 (announced jointly by then US president Bill Clinton and British Prime Minister Tony Blair on June 26, 2000). May 2006: sequence of the last chromosome for a total estimated cost of \$3 billion. The claims was/is that knowing the sequence of the genes would open immediately the door to great discoveries in medicine. 

However, for any reasonable scientist acquainted with complex systems (and the human body and its immune system in particular are clearly very complex systems by all definitions of the term), it was clear
that this was completely oversold: knowing the sequence of letters in a text without knowing the
language brings little and the major obstacles remain, which is that of understanding the text. This
involves many many-body problems (since hundreds to thousands genes act often in concert and each may have several actions in multitude of diseases and expressions). But this ``bubble''  or enthusiasm on genomics has served its goal of promoting a 
branch of research. This bubble led to huge investments, huge risks were taken,
and little return has occurred in most of the investment, except for some niches. It is interesting to read the statements of the community in the post-bubble era, emphasizing that decades are needed to really 
exploit the data.

\item {\bf Dolly the Sheep (July 5, 1996 Ð February 14, 2003).}  ``Suspendous'' and ``mind-boggling'' were just two reactions to the birth of Dolly, the first mammal cloned from an adult cell. Suddenly, the idea of herds of identical prize bulls, or sheep producing medicines for humans in their milk, seemed wholly plausible. Then there was therapeutic cloning, which would provide genetically matched human tissue to patch up even the most seriously ill patient...

A New Scientist article in 2006 
[Aldhous and Coghlan, 2006] states: ``Much of the excitement surrounding the creation of the first animal clone has vanished and therapeutic cloning is in the doldrums. But her influence should not be downplayed...'' A report in Nature [Check, 2007] confirms: ``...these advances haven't led to big improvements in the cloning process, or yielded huge commercial payoffs.''

\item {\bf The IT and Internet bubble until March 2000.}  The so-called ``new economy'' bubble on information technology and the internet has been associated with huge investments (and big losses) on the IT sector.  Few of the ``dot-com''
companies into which investors have poured billions of dollars of capital have
lived to pay them an adequate return.  Many companies have died but a few have survived and some have become giants (Yahoo, Google,...). 

The world has also set the pace on trying out new Internet business models, and presumably
will benefit enormously from the experience gained by the tens of thousands of
engineers, entrepreneurs, and web designers who have acquired human capital in a
new industry. Thus again, a bubble has left many ``dead'' but a few great innovations emerged, with the accumulation of new human capital.
\end{itemize}

I am sure many other examples will jump to the mind of the reader, that would warrant
a detailed discussion.

\section{Illusion of control}

Human beings like to believe they are in control of their destiny. This ubiquitous trait seems to increase motivation and persistence, and is probably evolutionarily adaptive. The success of science and technology, with the development of ever more sophisticated computerized integrated sensors in the biological, environmental and social sciences, illustrate the quest for control as a universal endeavor. The exercise of governmental authority, the managing of the economy, the regulation of financial markets, the management of corporations, and the attempt to master natural resources, control natural forces and influence environmental factors all arise from this quest. 

Langer [1975]'s phrase, ``illusion of control''  describes the fact that individuals appear hard-wired to over-attribute success to skill, and to underestimate the role of chance, when both are in fact present. Whether control is genuine or merely perceived is a prevalent question in psychological theories.
outcomes, especially when they are the result of aggregations of individual optimization processes. 
But how good really is our ability to control? How successful is our track record in these areas? There is little understanding of when and under what circumstances we may over-estimate or even lose our ability to control and optimize 

Satinover and Sornette [2007a,b] have demonstrated
analytically using the theory of Markov Chains and by numerical simulations in two classes of games, the Minority game and the Parrondo Games, that agents who optimize their strategy based on past information actually perform worse than non-optimizing agents. In other words, low-entropy (more informative) strategies under-perform high-entropy (or random) strategies. This provides a precise definition of the Òillusion of controlÓ in set-ups a priori defined to emphasize the importance of optimization

Our robust message is that, under bounded rationality, the simple (large-entropy) strategies are often to be preferred over more complex elaborated (low-entropy) strategies. This is a message that should appeal to managers and practitioners, who are well-aware in their everyday practice that simple solutions are preferable to complex ones, in the presence of the ubiquitous uncertainty.
More examples should be easy to find. For instance, control algorithms, which employ optimal parameter estimation based on past observations, have been shown to generate broad power law distributions of fluctuations and of their corresponding corrections in the control process, suggesting that, in certain situations, uncertainty and risk may be amplified by optimal control. In the same spirit, more quality control in code development often decreases the overall quality which itself spurs more quality control leading to a vicious circle. In finance,  there are many studies suggesting that most fund managers perform worse than random and strong evidence that over-trading leads to anomalously large financial volatility. Let us also mention the interesting experiments in which optimizing humans are found to perform worse than rats [Grandin and Johnson, 2004]. We conjecture that the illusion-of-control effect should be widespread in many strategic and optimization games and perhaps in many real life situations. This puts the question at a quantitative level so that it can be studied rigorously to help formulate better strategies and tools for management and policy, which take into account the intrinsic limitations of control in complex set-ups with feedbacks.

\section{Concluding remarks}

I have tried here to weave an outline of how the science of complexity, which include
evidence from both the natural and the social sciences, may help in policy
decision making. In this journey, as can be expected, we find more questions
than answers, more puzzles than solutions, more challenges than settlements.

Several other important ingredients have been left out. For instance, recent research
in cognitive sciences and in anthropology suggest that human beings are made
to interact with no more than about $150$ other humans [Dunbar, 1998], otherwise 
conflicts arise and fragmentation follows.
Actually, carefully studies of human groups show the existence
of a delicate hierarchy of natural group sizes [Zhou et al., 2005], that may be a result of
evolution. This suggests that the coordination of human activities in large modern
society needs to recognize this essential cognitive limitation probably deeply
rooted in our emotional and rational brain. 

Another ingredient which in my mind
needs to be incorporated in a science-based approach to policy 
decision making is the context-dependent, cultural as well as evolutionary
control of human cooperation, based on the existence of feedbacks such a
reward and punishments [Fehr and Fischbacher, 2003; Darcet and Sornette, 2006].
The fundamental question here is to identify the springs at the origin of cooperation
and the elements (structural and dynamical) that may hinder or destroy it.
 
\vskip 1cm

\noindent {\bf References}

Aldhous, P. and A. Coghlan, Dolly's cloning revolution fails to materialise, New Scientist, 2558, 01 July 2006, page 8.

Anifrani, J.-C.,  C. Le Floc'h, D. Sornette and B. Souillard,
 Universal Log-periodic correction to renormalization group scaling for rupture stress
prediction from acoustic emissions, J.Phys.I France 5 (6), 631-638 (1995).

Anifrani, J.-C.,  C. Le Floc'h and D. Sornette,
Pr\'ediction de la rupture de r\'eservoirs composites de haute pression \`a l'aide
de l'\'emission acoustique, Contr\^ole Industriel 220, 43-45 (1999).

 Bak, P., How Nature Works:
the Science of Self-organized Criticality (Copernicus,~New York, 1996).

Bak, P. and M. Paczuski, Complexity, contingency, and
criticality, Proc. Natl. Acad. Sci. USA, 92, 6689-6696 (1995).

Buchanan, M., Measuring Science Revolutions, Nature 384, 325 (1996).

Check, E., Cloning special: Dolly: a hart act to follow, Nature 445,
802 (22 February 2007) doi:10.1038/445802a.

Darcet, D. and D. Sornette,
Emergence of human cooperation and altruism by evolutionary feedback selection,
working paper, first version (2006)
(\url{http://arxiv.org/abs/physics/0610225})

Dieks and Chang, Differences in impact of scientific 
publications: some indices derived from a citation analysis, 
 Social Studies of Science 6, 247 (1976). 

Dunbar, R. I. M., The social brain hypothesis. Evol. Anthrop. 6, 178-190 (1998).

Ehrenreich, H., Strategic curiosity: Semiconductor physics in the 1950s,
Physics Today,  January, 28-34 (1995).

Einstein, A., \"Uber die von der molekularkinetischen Theorie der W\"arme
geforderte Bewegung von in ruhenden  Fl\"ussigkeiten suspendierten
Teilchen, Ann. Phys. 17, 549 (1905).

Einstein, A.,
Investigations on the Theory of Brownian Movement (New York: Dover, 1956).

Fehr, E. and Fischbacher, U. The nature of human altruism. Nature 425, 785-791 (2003).

Grabowski, H.G. and Vernon, J.M.,  A new look at the returns and
risks to pharmaceutical R\&D, Management Sci. 36, 804-821 (1990).

Grabowski, H.G. and Vernon, J.M., Returns to R\&D on new drug
introductions in the 1980s, J. Health Econ. 13, 383-406 (1994).

Grandin, T. and C. Johnson, Animals in translation, Scribner (2004).

Griffin, M., Space exploration: real reasons and acceptable reasons, 
Comments made at the Quasar Award Dinner, Bay Area
Houston Economic Partnership on 19 Jan 2007
(\url{http://www.spaceref.com/news/viewsr.html?pid=23189})

Kuhn, T.,  The Structure of Scientific Revolutions, 2nd ed.,  
Chicago: University of Chicago Press (1970). 

Langer, E.J., Journal of Personality and Social Psychology 32 (2), 311-328 (1975).

 Le Floc'h, C. and D. Sornette,
Predictive acoustic emission: Application on helium high pressure tanks,
Pr\'ediction des \'ev\`enements catastrophiques: une 
nouvelle approche pour le controle de sant\'e structurale,
Instrumentation Mesure Metrologie (published by Hermes Science)
RS serie I2M volume 3 (1-2), 89-97 (2003).
  
Martin, Ben R., et al., The relationship between publicly funded
basic research and economic performance, Report of the Science Policy Research Unit
of the University of Sussex (1996). 

Maveyraud, C., J.P. Vila, D. Sornette, C.Le FlocÕh, J.M. Dupillier, R. Salom\'e,
Numerical modelling of the behaviour of high-pressure vessels under a hypervelocity impact,
Mec.Ind. 2, 57Ð62 (2001).
 
National Academy of Sciences, National
Academy of Engineering, Institute of Medecine and National Research Council, 
Preparing for the 21st Century (1997) (\url{http://www2.nas.edu/21st/})

Redner, S., How popular is your paper? An empirical study of the citation. distribution,
Euro. Phys. J. B 4, 131-134 (1998).

Reichhardt, T., A measure of happiness, Nature 444, 418-419 (2006)

Romer, D., {\it Advanced macroeconomics} (McGraw-Hill, New York, 1996).

Ruelle, D., Conversations on nonequilibrium physics with 
an extraterrestrial,  Physics Today, 57(5), 48-53 (2004).

Satinover, J.B. and D. Sornette, 
``Illusion of Control'' in Minority and Parrondo Games,
submitted to theEuropean Journal of Physics B (2007a)
(\url{http://arxiv.org/abs/0704.1120})

Satinover, J.B. and D. Sornette, 
Illusion of Control in a Brownian Game, submitted to
Physica A (2007b)
(\url{http://arxiv.org/abs/physics/0703048})

Scherer, F.M., The size distribution of profits from innovation, Annale dÕEconomie et de Statistique 49/50, 495-516 (1998).

Schumpeter, J.A., 
{\it Business Cycles: A Theoretical, Historical and Statistical Analysis of the
Capitalist Process} (McGraw-Hill, New York, 1939).

Sornette, D. Economy of scales in innovations with block-busters, Quantitative Finance 2, 224Ð227 (2002a).

Sornette, D.,  Predictability of catastrophic events: material rupture, earthquakes, turbulence, 
financial crashes and human birth, Proc. Natl. Acad.
Sci. USA, 99 SUPP1, 2522-2529 (2002b).

Sornette, D., Why Stock Markets Crash (Critical Events in Complex Financial Systems)
(Princeton University Press, Princeton, NJ, 2003a).

Sornette, D., Critical market crashes, Physics Reports 378 (1), 1-98 (2003b).

Sornette, D., Critical Phenomena in Natural Sciences,
Chaos, Fractals, Self-organization and Disorder: Concepts and Tools,
2nd ed. (Springer Series in Synergetics, Heidelberg, 2004).

Sornette, D., Endogenous versus exogenous origins of crises,
in the monograph entitled ``Extreme Events in Nature and Society,'' 
Series:  The Frontiers Collection  S. Albeverio, V. Jentsch and H. Kantz, eds.  (Springer, Heidelberg, 2005) (\url{http://arxiv.org/abs/physics/0412026})

Sornette, D. and D. Zajdenweber,
The economic return of research\,: the Pareto law and its implications, 
European Physical Journal B, 8 (4), 653-664 (1999).

Stratonovich,~R.L. ,
Nonlinear Nonequilibrium Thermodynamics I: Linear and Nonlinear
Fluctuation-Dissipation Theorems (Springer,~Berlin; New York, 1992).

Zhou,  W.-X., D. Sornette, R.A. Hill and R.I.M. Dunbar,
Discrete Hierarchical Organization of Social Group Sizes, 
Proc. Royal Soc. London 272, 439-444 (2005).

\end{document}